\newcommand{\diracslash}[1]{#1\llap{/\kern2pt}}

\newcommand{\be}{\begin{equation}}
\newcommand{\ee}{\end{equation}}
\newcommand{\bea}{\begin{eqnarray}}
\newcommand{\eea}{\end{eqnarray}}
\newcommand{\ba}[1]{\begin{array}{#1}}
\newcommand{\ea}{\end{array}}

\newcommand{\bt}{\begin{tabular}}
\newcommand{\et}{\end{tabular}}

\newcommand{\beas}{\begin{eqnarray*}}
\newcommand{\eeas}{\end{eqnarray*}}

\documentclass[preprint,prd,aps,floats,nofootinbib,floatfix]{revtex4}
\usepackage{graphicx}
\begin{document}

\title{Charmonium states in strong magnetic fields} 
\author{Amal Jahan CS}
\email{amaljahan@gmail.com}
\affiliation{Department of Physics, Indian Institute of Technology, Delhi,
Hauz Khas, New Delhi -- 110 016, India}

\author{Nikhil Dhale}
\email{dhalenikhil07@gmail.com}
\affiliation{Department of Physics, Indian Institute of Technology, Delhi,
Hauz Khas, New Delhi -- 110 016, India}

\author{Sushruth Reddy P}
\email{sushruth.p7@gmail.com}
\affiliation{Department of Physics, Indian Institute of Technology, Delhi,
Hauz Khas, New Delhi -- 110 016, India}

\author{Shivam Kesarwani}
\email{kesarishvam@gmail.com}
\affiliation{Department of Physics, Indian Institute of Technology, Delhi,
Hauz Khas, New Delhi -- 110 016, India}

\author{Amruta Mishra}
\email{amruta@physics.iitd.ac.in}
\affiliation{Department of Physics, Indian Institute of Technology, Delhi,
Hauz Khas, New Delhi -- 110 016, India}

\begin{abstract}
The medium modifications of the masses of the charmonium states
(J/$\psi$, $\psi$(3686) and $\psi$(3770)) in asymmetric nuclear matter
in the presence of strong magnetic fields are studied using an
effective chiral model. The mass modifications arise due to
medium modifications of the scalar dilaton field, which simulates
the gluon condensates of QCD within the effective hadronic model. 
The effects due to the magnetic field as well as
isospin asymmetry are observed to be appreciable
at high densities for the charmonium states, which can have
consequences, e,g, in the production of the open charm mesons
and the charmonium states, in the asymmetric heavy ion collisions 
at the  compressed baryonic matter (CBM) experiments
at the future facility at GSI. The presence of magnetic field
leads to Landau quantization of the energy levels of the proton
in the asymmetric nuclear matter. The  effects of the anomalous
magnetic moments of the proton and neutron on the masses
of the charmonium states are studied and are observed to lead to
larger masses of the charmonium states, as compared to when 
these effects are not taken into account.
\end{abstract}
\maketitle

\def\bfm#1{\mbox{\boldmath $#1$}}

\section{Introduction}

The study of hadrons under extreme conditions, e.g., at high
densities and/or temperatures is a topic of intense research
in strong interaction physics, due to its relevance in the 
context of ultra relativistic heavy ion collision experiments.
In the recent past, the study of medium modifications of the
heavy flavour hadrons have also gained considerable interest 
\cite{Hosaka_Prog_Part_Nucl_Phys} 
because of the possibility of formation of the heavy hadrons
bound in nuclei due to their attractive interaction in nuclear matter.  
The medium modifications of the heavy flavour mesons are also
of relevance to the experimental observables, e.g., the production 
and propagation of these mesons 
in the strongly interacting matter resulting from high energy 
nuclear collisions. Large magnetic fields are believed to be
created in these relativistic heavy ion collision experiments
\cite{HIC_mag}
and the time evolution of the magnetic field
\cite{time_evolution_B_HIC_Tuchin,time_evolution_B_HIC_Ajit}
is still an open problem, which needs the proper estimate
of the electrical conductivity of the medium as well as 
solutions of the magnetohydrodynamic equations.
The estimation of the magnetic fields being ceated
in non-central ultra-relativistic heavy-ion collision
experiments as huge, has initiated the study of 
the heavy flavour mesons,
e.g. the open charm and open bottom mesons
\cite{Gubler_D_mag_QSR,machado_1,B_mag_QSR,dmeson_mag,bmeson_mag}
as well as the charmonium states
\cite{charmonium_mag_QSR,charmonium_mag_lee} 
in the presence of strong magnetic fields. 
The isospin asymmetry effects 
also need to be accounted for the study of the in-medium
properties of the hadrons due to the fact that the 
colliding nuclei in the heavy ion collision experiments
have large isospin asymmetry. In the present work, we study
the medium modifications of the charmonium states (ground state 
$J/\psi$ and excited states $\psi(3686)$ and $\psi(3770)$)
in isospin asymmetric nuclear matter in the presence of
strong magnetic fields using a chiral effective model. 

The heavy quarkonium states, which are bound states of heavy
quark, $Q=(c,b)$ and heavy antiquark, $\bar Q=(\bar c, \bar b)$,
have been  studied extensively in the literature 
using the potential models \cite{eichten,satz,repko}.
In these models, the mass of the quarkonium state is obtained 
by solving the Schrodinger equation using 
an effective potential. The Cornell potential, which has been
widely used in the literature to describe the charmonium and 
bottomonium spectroscopy, has a short distance Coulomb type 
of interaction along with a long distance confining potential.
At finite temperatures, there are color Debye screening 
effects, which can lead to dissociation of the quarkonium state
at a critical temperature, $T_c$. 
In Refs. \cite{pes1,pes2,voloshin},
the heavy quarkonium state has been studied 
as a non-relativistic bound state of the heavy quark ($Q$) 
and heavy antiquark ($\bar Q$) interacting by the color Coulomb 
potential and 
the effects of the gluonic fluctuations on the quarkonium state
have been considered in the vacuum/nuclear medium. With the
assumption that the separation of $Q$ and $\bar Q$
in the  quarkonium bound state is small compared to 
the characteristic scale of the gluonic fluctuations, 
the interaction of the quarkonium state with the gluonic field
is expanded in multipole expansion. The leading contribution
to the mass of the quarkonium state is proportional 
to the gluon condensates in vacuum/nuclear medium.
In the QCD sum rule calculations, the mass modifications 
of the heavy quarkonium states,
e.g., the charmonium states arise due to the medium modifications 
of the gluon condensate in the nuclear medium 
\cite{kimlee,klingl,amarvjpsi_qsr}. 
This is contrary to the light vector mesons ($\omega$, $\rho$
and $\phi$) which are modified in the hadronic medium, due to 
the medium changes in the light quark condensates
in the hadronic medium \cite{hatsuda,am_vecmeson_qsr}, 
within the QCD sum rule approach. 
The open heavy flavour mesons, which 
consist of a light quark (antiquark) along with a heavy
(charm or bottom) antiquark (quark) are also modified due to the
interaction with the light quark condensates in the nuclear
medium within the framework of QCD sum rule approach 
\cite{open_heavy_flavour_qsr,Wang_heavy_mesons,arvind_heavy_mesons_QSR}. 
The open heavy flavour mesons have also been studied in the literature,
using quark meson coupling model \cite{open_heavy_flavour_qmc},
where the quarks interact with exchange of scalar and vector mesons
\cite{qmc}, using heavy quark symmetry and interaction
of these mesons with nucleons via pion exchange \cite{Yasui_Sudoh_pion},
heavy meson effective theory
\cite{Yasui_Sudoh_heavy_meson_Eff_th}, heavy flavour meson as
an impurity in nuclear matter \cite{Yasui_Sudoh_heavy_particle_impurity}
as well as using the coupled channel approach 
\cite{ltolos,ljhs,mizutani,HL,tolos_heavy_mesons}. 
In Ref. \cite{leeko}, the mass modifications of the charmonium states 
have been studied using leading order QCD formula \cite{pes1} 
and the linear density approximation for
the gluon condensate in the nuclear medium. 
The QCD sum rule calculations for the charmonium states 
have been  generalized to finite temperatures \cite{moritalee},
where the medium modifications of these states have been
studied arising due to the temperature effects on the gluon 
condensates, extracted from lattice calculations.
Within the chiral effective model \cite{Schechter,paper3,kristof1}
used in the present investigation, the gluon condensate is mimicked 
by incorporating a scalar dilaton field.
The medium modifications of the charmonium masses are
obtained from the modifications of the dilaton field 
\cite{amarindamprc,amarvdmesonTprc,amarvepja}
within the chiral effective model.
Due to the attractive interaction of the $J/\psi$ 
in nuclear matter \cite{leeko,amarvdmesonTprc,amarvepja,krein_jpsi},
the possibility of the $J/\psi$ forming bound states 
with nuclei have also been predicted \cite{krein_17}.

In the present work, we investigate the medium modifications of the masses 
of the charmonium states in the isospin asymmetric nuclear medium
in the presence of strong magnetic fields using a chiral effective model. 
These are studied from the medium modifications of the scalar
dilaton field introduced in the model to incorporate
the broken scale invariance of QCD. The dilaton field is
thus related to the gluon condensates of QCD and its medium
modification gives the measure for the medium modification
of the gluon condensate, which is used to calculate the in-medium
masses of the charmonium states.
The chiral effective model has been used extensively in the literature,
for the study of finite nuclei \cite{paper3},
strange hadronic matter \cite{kristof1}, 
light vector mesons \cite{hartree}, 
strange pseudoscalar mesons, e.g. the kaons and antikaons
\cite{kaon_antikaon,isoamss,isoamss1,isoamss2}
in isospin asymmetric hadronic matter,
as well as for the study of bulk matter of neutron stars 
\cite{pneutronstar}.
The model has also been generalized to include the charm and bottom 
sectors, to study the open (strange) charm mesons
\cite{amarindamprc,amarvdmesonTprc,amarvepja,amdmeson,DP_AM_Ds},
open (strange) bottom mesons \cite{DP_AM_bbar,DP_AM_Bs}, 
the charmonium states \cite{amarvdmesonTprc,amarvepja}, 
the upsilon states \cite{AM_DP_upsilon}, the partial decay widths
of the charmonium states to $D\bar D$ in the hadronic medium
\cite{amarvepja} using a light quark creation model \cite{friman},
namely the $^3P_0$ model \cite{3p0}. The in-medium
decay widths of the charmonium (bottomonium) to $D\bar D$
($B\bar B$) have also been investigated using a field theoretic model
for composite hadrons \cite{amspmwg,amspm_upsilon}.
The model has recently been used to study the in-medium masses
of the open charm ($D$ and $\bar D$ mesons) \cite{dmeson_mag}
as well as the open bottom ($B$ and $\bar B$ mesons) \cite{bmeson_mag}
in strongly magnetized nuclear matter. 

The outline of the paper is as follows : In section II, we describe
briefly the chiral effective model used to study the in-medium 
masses of the charmonium states in isospin asymmetric nuclear matter
in the presence of an external magnetic field. The medium 
modifications of the charmonium masses arise from the medium 
modification of a scalar dilaton field introduced in the hadronic 
model to incorporate broken scale invariance of QCD leading to QCD 
trace anomaly. The effects of the anomalous magnetic moments of the
nucleons are also taken into account in the present work. 
In section III, we discuss the results obtained in the present
investigation of the in-medium masses of the charmonium states
in strong magnetic fields and section IV summarizes the findings of 
the present study.

\section{In-medium masses of the charmonium states}

We use an effective chiral $SU(3)$ model \cite{paper3} 
to study the in-medium masses
of the charmonium states in asymmetric nuclear matter
 in the presence of an external magnetic field.
The model is based on the nonlinear realization of chiral 
symmetry \cite{weinberg, coleman, bardeen} and broken scale invariance 
\cite{paper3,kristof1,hartree}. 
The effective hadronic chiral Lagrangian density contains the following terms
\be
{\cal L} = {\cal L}_{kin} + \sum_{ W =X,Y,V,{\cal A},u }{\cal L}_{BW}
          + {\cal L}_{vec} + {\cal L}_0 +
{\cal L}_{scalebreak}+ {\cal L}_{SB}+{\cal L}_{mag},
\label{genlag} \ee 
In the above Lagrangian density, the first term ${\cal L}_{kin}$ 
corresponds to the kinetic energy terms of the baryons and the mesons.
${\cal L}_{BW}$ is the baryon-meson interaction term and the
the vacuum baryon masses are generated by the 
baryon-scalar meson interaction terms.
${\cal L}_{vec}$  describes the interactions of the vector 
mesons, ${\cal L}_{0}$ contains 
the meson-meson interaction terms inducing the spontaneous breaking of 
chiral symmetry. 
${\cal L}_{scalebreak}$ is a scale invariance breaking logarithmic 
potential given in terms of a scalar dilaton field \cite{heide1}, 
$ {\cal L}_{SB} $ describes the explicit chiral symmetry
breaking, and ${\cal L}_{mag}$ is the contribution 
from the magnetic field, given as \cite{broderick1,broderick2,Wei,mao}
\be 
{\cal L}_{mag}=-{\bar {\psi_i}}q_i \gamma_\mu A^\mu \psi_i
-\frac {1}{4} \kappa_i \mu_N {\bar {\psi_i}} \sigma ^{\mu \nu}F_{\mu \nu}
\psi_i
-\frac{1}{4} F^{\mu \nu} F_{\mu \nu},
\label{lmag}
\ee
where, $\psi_i$ corresponds to the $i$-th baryon (proton and neutron 
for nuclear matter, as considered in the present work).
The second term in equation (\ref{lmag}) corresponds 
to the tensorial interaction
with the electromagnetic field and is related to the
anomalous magnetic moment 
\cite{broderick1,broderick2,Wei,mao,amm,VD_SS,aguirre_fermion}
of the nucleon.

The concept of broken scale invariance of QCD
is simulated in the effective 
Lagrangian at tree level \cite{Schechter} through the introduction of 
the scale breaking term \cite{paper3,kristof1}
of the general Lagrangian given by equation ({\ref{genlag}) as
\begin{equation}
{\cal L}_{scalebreak} =  -\frac{1}{4} \chi^{4} {\rm {ln}}
\Bigg ( \frac{\chi^{4}} {\chi_{0}^{4}} \Bigg ) + \frac{d}{3}{\chi ^4} 
{\rm {ln}} \Bigg ( \bigg (\frac{I_{3}}{{\rm {det}}\langle X 
\rangle _0} \bigg ) \Bigg ),
\label{scalebreak}
\end{equation}
where $I_3={\rm {det}}\langle X \rangle$, with $X$ as the multiplet
for the scalar mesons. 
The effect of these logarithmic terms is to break the scale invariance, 
which leads to the trace of the energy momentum tensor as \cite{paper3}
\begin{equation}
\theta_{\mu}^{\mu} = \chi \frac{\partial {\cal L}}{\partial \chi} 
- 4{\cal L} 
=-\chi^{4}.
\label{tensor1}
\end{equation}
The comparison of the trace of the energy momentum tensor of QCD
in the massless quarks limit \cite{cohen} to that of the chiral 
effective model as used in the present work,
leads to the relation of the dilaton field to the scalar gluon condensate
as given by \cite{heide1}
\begin{equation}
\Theta_{\mu}^{\mu} = \langle \frac{\beta_{QCD}}{2g} 
G_{\mu\nu}^{a} G^{\mu\nu a} \rangle  \equiv  -(1-d)\chi^{4} 
\label{tensor2}
\end{equation}
The parameter $d$ originates from the second logarithmic term of equation 
(\ref{scalebreak}). 
The QCD $\beta$ function at one loop level, for 
$N_{c}$ colors and $N_{f}$ flavors, is given by
\begin{equation}
\beta_{\rm {QCD}} \left( g \right) = -\frac{11 N_{c} g^{3}}{48 \pi^{2}} 
\left( 1 - \frac{2 N_{f}}{11 N_{c}} \right),  
\label{beta}
\end{equation}
The first term in the above equation is a purely gluonic contribution,
whereas the second term arises from the screening contribution
of the quark pairs. At the one loop level of the $\beta$ function,
the value of $d$ should be $\frac{2 N_f}{11 N_c}$ as  is evident
from equations (\ref{tensor2}) and (\ref{beta}). However, as one
can not rely on the one loop level result, the parameter $d$ is
taken as a parameter within the chiral effective model \cite{paper3}
used in the present investigation.

The trace of the energy-momentum tensor in QCD, using the 
one loop beta function given by equation (\ref{beta}),
for $N_c$=3 and $N_f$=3, is given as,
\begin{equation}
\theta_{\mu}^{\mu} = - \frac{9}{8} \frac{\alpha_{s}}{\pi} 
G_{\mu\nu}^{a} G^{\mu\nu a}
\label{tensor4}
\end{equation} 
Using equations (\ref{tensor2}) and (\ref{tensor4}), we can write  
\begin{equation}
\left\langle  \frac{\alpha_{s}}{\pi} G_{\mu\nu}^{a} G^{ \mu\nu a} 
\right\rangle =  \frac{8}{9}(1 - d) \chi^{4}
\label{chiglu}
\end{equation}
We thus see from the equation (\ref{chiglu}) that the scalar 
gluon condensate $\left\langle \frac{\alpha_{s}}{\pi} G_{\mu\nu}^{a} 
G^{\mu\nu a}\right\rangle$ is proportional to the fourth power of the 
dilaton field, $\chi$, in the chiral SU(3) model.
As mentioned earlier, the masses of the charmonium states are 
modified in the nuclear medium due to the modifications of the
gluon condensates, which is calculated within the chiral SU(3) model,
from the modification of the dilaton in the medium,
by using the equation (\ref{chiglu}). 
In the present work of study of the in-medium charmonium masses
using the chiral SU(3) model, we use the mean field approximation,
where all the meson fields are treated as classical fields. 
In this approximation, only the scalar and the vector fields 
contribute to the baryon-meson interaction, ${\cal L}_{BW}$
since for all the other mesons, the expectation values are zero.
In the present investigation, for given values of the external
magnetic field, the scalar-isoscalar fields
(non strange, $\sigma$ and strange, $\zeta$ fields), the 
scalar isovector field $\delta$, and the dilaton field,
$\chi$ are solved from the coupled equations of motion,
for given values of the baryon density, $\rho_B$, and the isospin 
asymmetry parameter, $\eta= ({\rho_n -\rho_p})/({2 \rho_B})$,
where $\rho_n$ and $\rho_p$ are the number densities of the neutron
and the proton respectively.
The mass modifications of the charmonium states ($J/\psi$,
$\psi(3686)$ and $\psi(3770)$), as arising from the medium modifications
of the dilaton field, $\chi$, in hot asymmetric nuclear (hyperonic) matter
\cite{amarvdmesonTprc,amarvepja}
have already been studied using the chiral SU(3) model. 
The mass shift of the charmonium states in the medium due to 
the gluon condensates \cite{leeko} in the large charm mass limit
has been used to calculate the mass shift of the charmonium states.
The medium change of the scalar gluon condensates,
which in the nonrelativistic limit, is due to the medium change
in the $\langle \frac{\alpha_s}{\pi} {\vec E}^2 \rangle$
\cite{leeko}, similar to the second order Stark effect.
In Ref. \cite{leeko}, the medium modifications of the
scalar gluon condensate was obtained in the linear density 
approximation. In the chiral SU(3) model as used in the
present investigation, the mass shift in the charmonium
state arises due to the medium modification of the
scalar gluon condensate, and hence due to the change 
in the value of the dilaton field, and is given as
\cite{amarvdmesonTprc,amarvepja}
\begin{equation}
\Delta m_{\psi}= \frac{4}{81} (1 - d) \int dk^{2} 
\langle \vert \frac{\partial \psi (\vec k)}{\partial {\vec k}} 
\vert^{2} \rangle
\frac{k}{k^{2} / m_{c} + \epsilon}  \left( \chi^{4} - {\chi_0}^{4}\right), 
\label{masspsi}
\end{equation}
where 
\begin{equation}
\langle \vert \frac{\partial \psi (\vec k)}{\partial {\vec k}} 
\vert^{2} \rangle
=\frac {1}{4\pi}\int 
\vert \frac{\partial \psi (\vec k)}{\partial {\vec k}} \vert^{2}
d\Omega,
\end{equation}
$m_c$ is the mass of the charm quark, taken as 1.95 GeV,
$m_\psi$ is the vacuum mass of the charmonium state 
and $\epsilon = 2 m_{c} - m_{\psi}$ is the binding energy. 
The formula for the mass shift of the heavy quarkonium state
given by equation (\ref{masspsi}) is derived assuming
Coulomb potential
between $Q$ and $\bar Q$ in the quarkonium state.
The value of $m_c$ as 1.95 GeV is taken so as to reproduce 
the mass difference of the charmonium states
$J/\psi$ and $\psi(3686)$ in vacuum \cite{leeko},
For $2m_c < m_\psi$, the binding energy becomes
negative, when the charmonium state  can no longer
exist as a bound state.
The values of the dilaton field
in the nuclear medium and the vacuum are $\chi$ and $\chi_0$  
respectively.
$\psi (k)$ is the wave function of the charmonium state
in the momentum space, normalized as $\int\frac{d^{3}k}{(2\pi)^{3}} 
\vert \psi(k) \vert^{2} = 1 $ \cite{moritalee}.

The wave functions for the charmonium states 
are taken to be Gaussian and are given as \cite{friman}
\begin{equation}
\psi_{N, l} = N\times Y_{l}^{m} (\theta, \phi) 
(\beta^{2} r^{2})^{\frac{1}2{} l} exp\Big({-\frac{1}{2} \beta^{2} r^{2}}
\Big) 
L_{N - 1}^{l + \frac{1}{2}} \left( \beta^{2} r^{2}\right)
\label{wavefn} 
\end{equation} 
where $\beta^{2} = M \omega / h$ characterizes the strength of the 
harmonic potential, $M = m_{c}/2$ is the reduced mass of 
the charm quark and charm anti-quark system, and $L_{p}^{k} (z)$ 
is the associated Laguerre Polynomial. As in Ref. \cite{leeko},
the oscillator constant $\beta$ is determined from the mean squared 
radii $\langle r^{2} \rangle$ as 0.46$^{2}$ fm$^2$, 0.96$^{2}$ fm$^2$ 
and 1 fm$^{2}$ for the charmonium states $J/\psi(3097) $, $\psi(3686)$ and 
$\psi(3770)$, respectively. This gives the value for the parameter
$\beta$ (in GeV) as 0.51, 0.38 and 0.37 for $J/\psi(3097)$, 
$\psi(3686)$ and $\psi(3770)$, assuming that these 
charmonium states are in the 1S, 2S and 1D states respectively. 
Knowing the wave functions of the charmonium states and 
calculating the medium modification of the dilaton field
in the magnetized nuclear matter, we obtain the mass shifts of the
charmonium states, $J/\psi$, $\psi (3686)$ and $\psi (3770)$
respectively. In the next section we shall present the results 
obtained for the in-medium charmonium masses 
in asymmetric nuclear matter in presence of strong magnetic fields.

\section{Results and Discussions}

In this section, we first investigate the effects of
magnetic field, density and isospin asymmetry of the magnetized
nuclear medium on the dilaton field $\chi$, which mimicks
the gluon condensates of QCD, within the chiral SU(3) 
model. The in-medium masses of charmonium states $J/\psi$, 
$\psi(3686)$ and $\psi(3770)$ are then calculated from 
the value of $\chi$ in the nuclear medium using 
equation (\ref{masspsi}). 

\begin{figure}
\includegraphics[width=16cm,height=16cm]{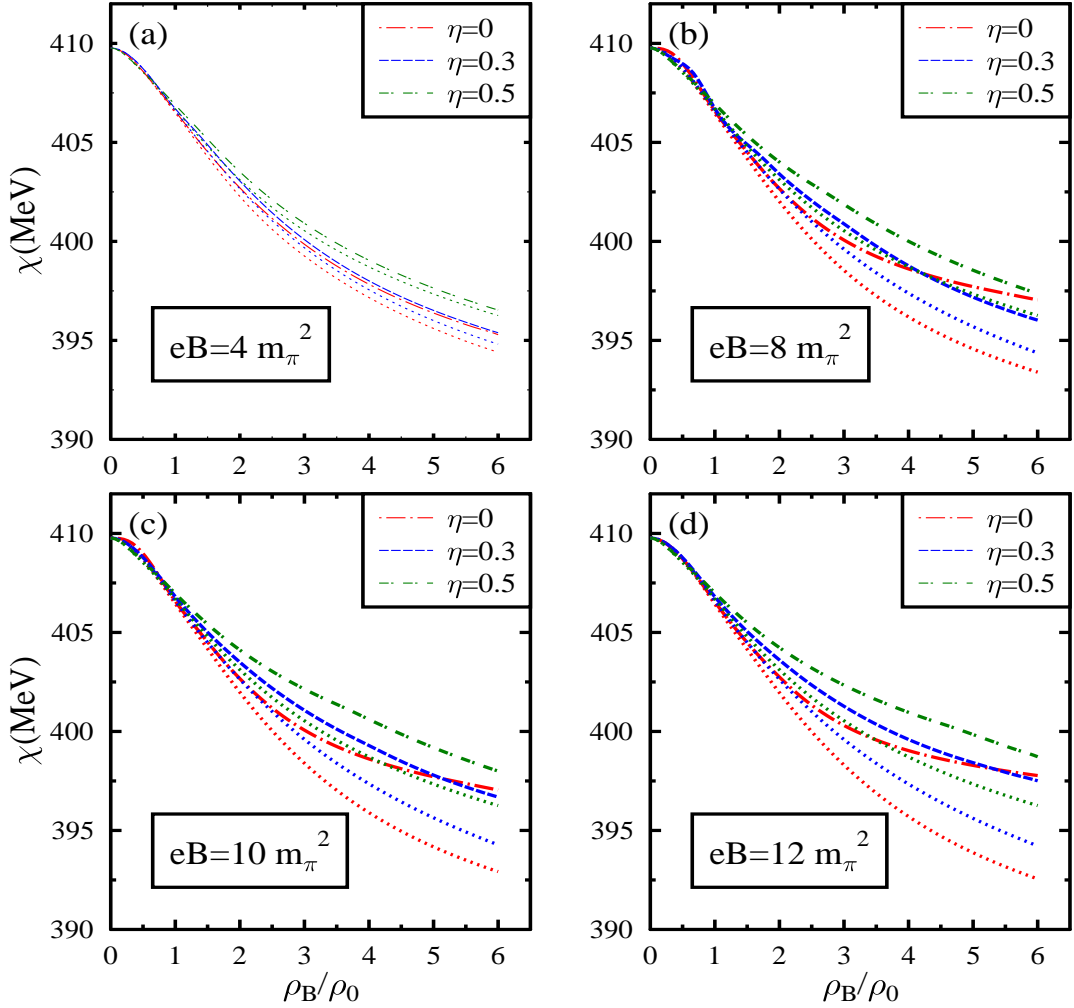}
\caption{(Color online)
The dilaton field $\chi$ plotted as a function of the
baryon density in units of nuclear matter saturation density,
for different values of the magnetic field and isospin 
asymmetry parameter, $\eta$, including the effects of the
anomalous magnetic moments of the nucleons. The results
are compared to the case when the effects of anomalous magnetic 
moments are not taken into consideration (shown as dotted lines).
}
\label{chi_mag}
\end{figure}

\begin{figure}
\includegraphics[width=16cm,height=16cm]{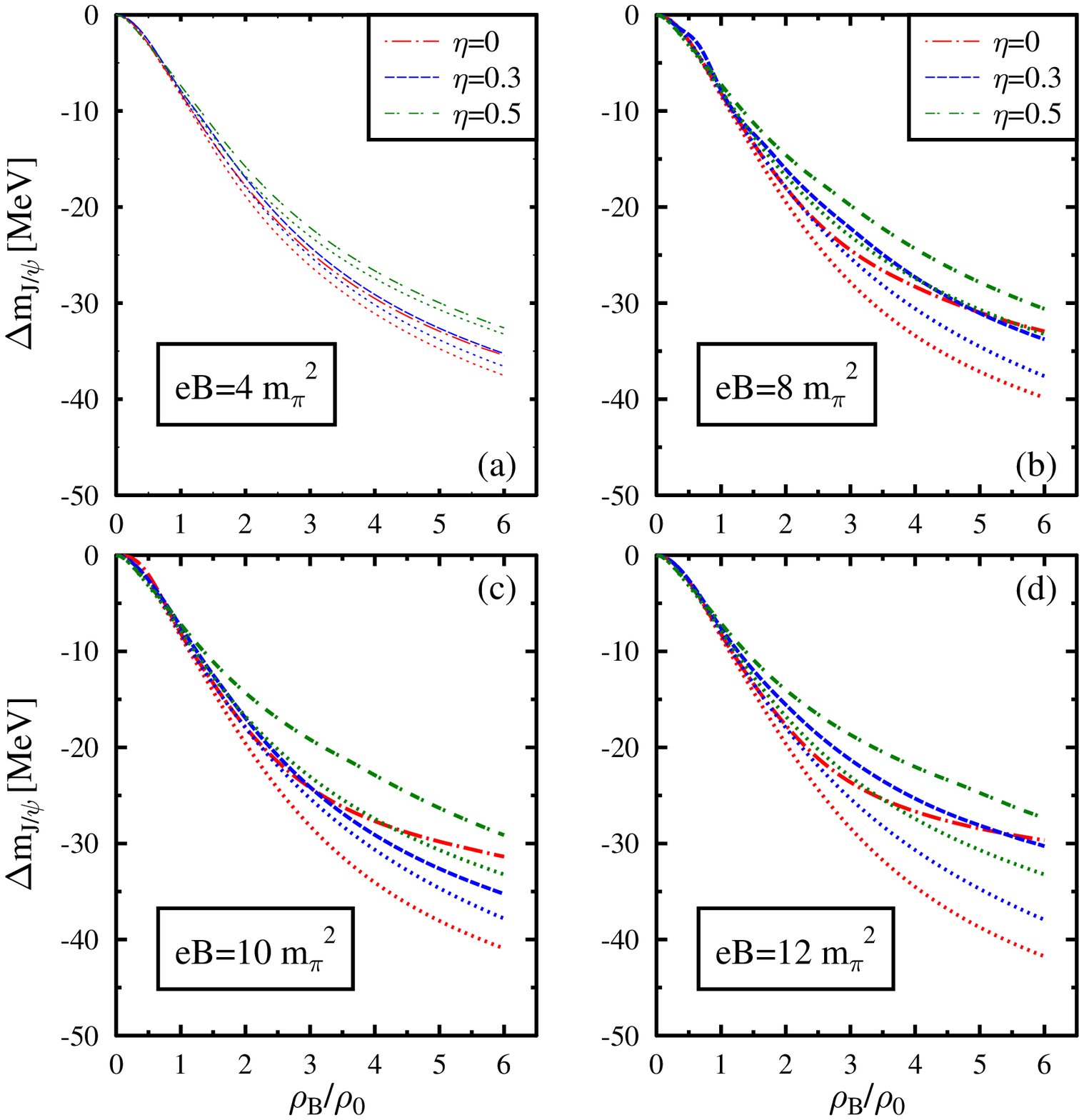}
\caption{(Color online)
The mass shift of J/$\psi$ plotted as a function of the
baryon density in units of nuclear matter saturation density,
for different values of the magnetic field and isospin 
asymmetry parameter, $\eta$, including the effects of the
anomalous magnetic moments of the nucleons. The results
are compared to the case when the effects of anomalous magnetic 
moments are not taken into consideration (shown as dotted lines).
}
\label{mjpsi_mag}
\end{figure}

\begin{figure}
\includegraphics[width=16cm,height=16cm]{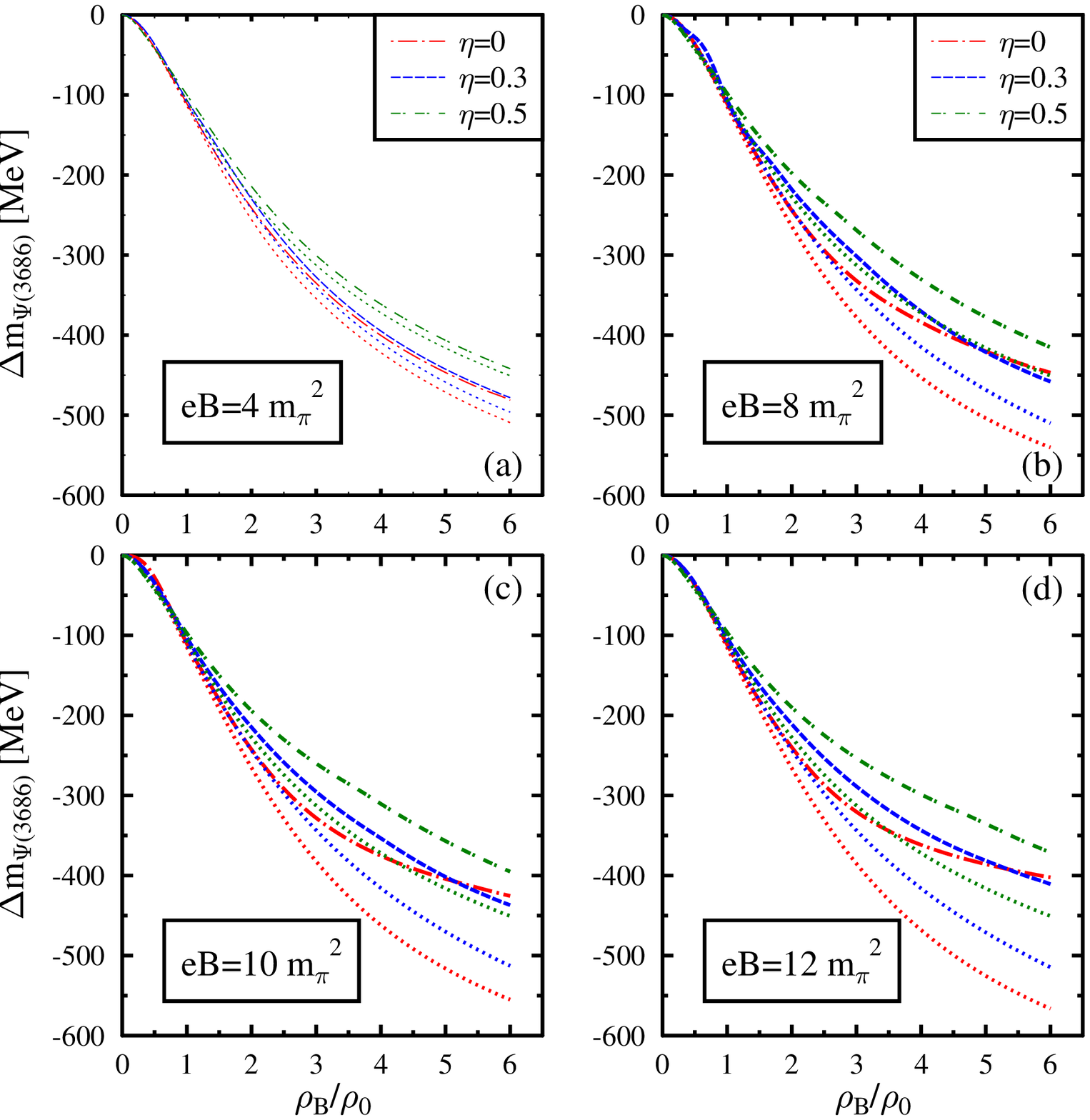}
\caption{(Color online)
The mass shift of $\Psi (3686)$ plotted as a function of the
baryon density in units of nuclear matter saturation density,
for different values of the magnetic field and isospin 
asymmetry parameter, $\eta$, including the effects of the
anomalous magnetic moments of the nucleons. The results
are compared to the case when the effects of anomalous magnetic 
moments are not taken into consideration (shown as dotted lines).
}
\label{mpsi_2s_mag}
\end{figure}

\begin{figure}
\includegraphics[width=16cm,height=16cm]{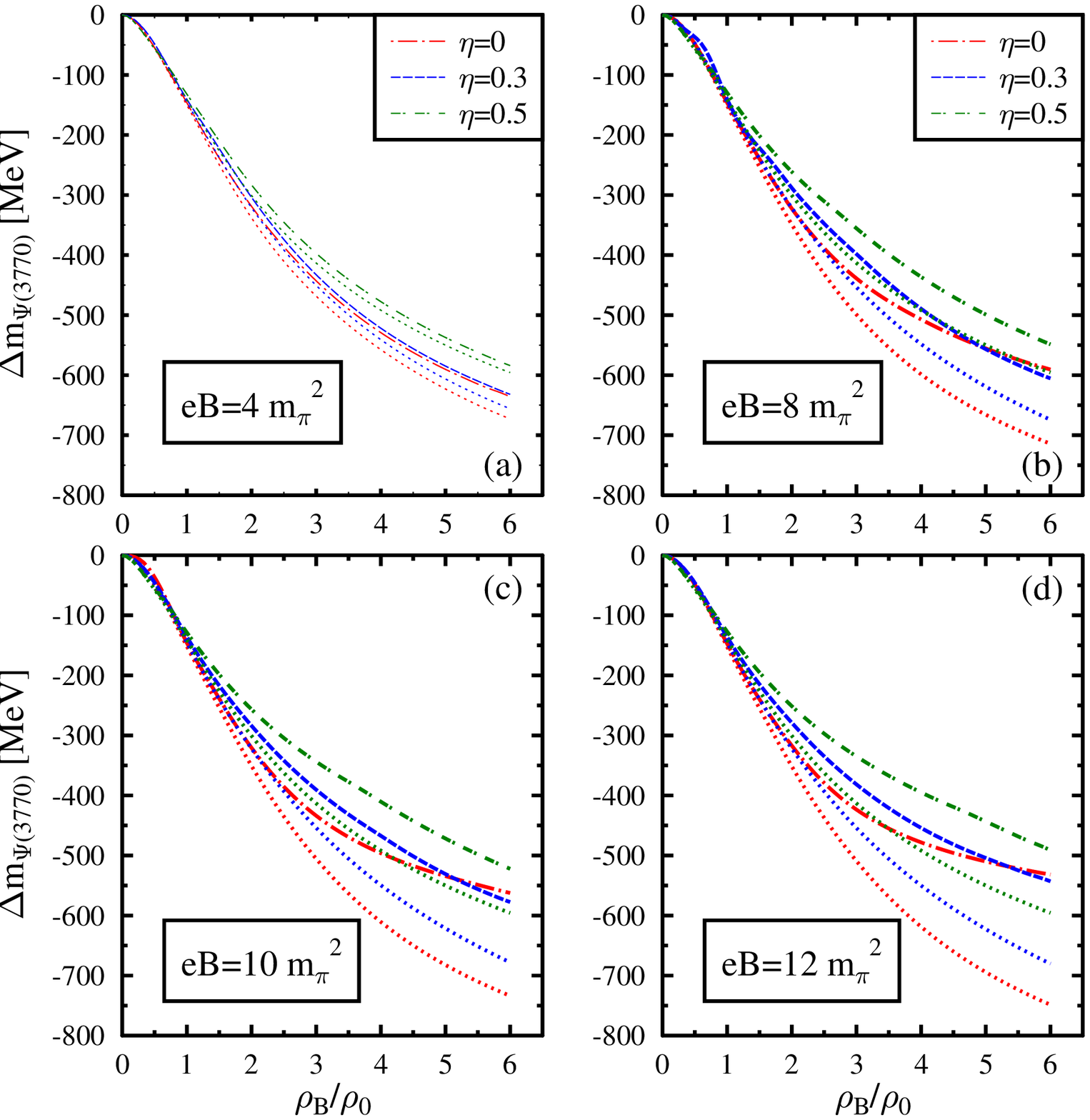}
\caption{(Color online)
The mass shift of $\Psi (3770)$ plotted as a function of the
baryon density in units of nuclear matter saturation density,
for different values of the magnetic field and isospin 
asymmetry parameter, $\eta$, including the effects of the
anomalous magnetic moments of the nucleons. The results
are compared to the case when the effects of anomalous magnetic 
moments are not taken into consideration (shown as dotted lines).
}
\label{mpsi_1d_mag}
\end{figure}

The dilaton field $\chi$ as modified
in the asymmetric nuclear medium in the presence of strong magnetic fields,
is plotted in figure \ref{chi_mag}.  The graph shows 
the behaviour of the dilaton field $\chi$ with density, for given values
of magnetic field and isospin asymmetry parameter, $\eta$,
accounting for the anomalous magnetic moments (AMM) of the nucleons and
compared with the situations when AMMs are not taken into consideration.
The variations of the dilaton field $\chi$ with magnetic field,
baryon density, and isospin asymmetry, within the chiral SU(3) model, 
are obtained by solving the coupled equations of motion of the
scalar fields, $\sigma$, $\zeta$, $\delta$ and $\chi$. 
These equations for the scalar fields contain the number and scalar
densities of the nucleons, where the effects of the magnetic field,
are in terms of the summation over the Landau energy levels
(for proton, the charged nucleon)
and the anomalous magnetic moments of the nucleons.
It can be observed that with increase in density, the value of $\chi$ 
decreases and the shift from its vacuum value ($\chi_0$= 409.77 MeV)
increases with density. 
With increasing magnetic field, for a certain density, $\chi$ is
observed to attain a higher value and thus 
the shift from the vacuum value decreases. 
When the anomalous magnetic moments of the nucleons are ignored, 
these deviations from vacuum value, are observed to increase 
with increasing magnetic field. 
For the isospin symmetric nuclear 
matter ($\eta$=0), for baryon density, $\rho_B$=$\rho_0$,
the nuclear matter saturation density, 
and for values of the external
magnetic field with $|eB|$, as $4m_\pi^2$, 
$8m_\pi^2$, $10m_\pi^2$, $12m_\pi^2$,
shown in panels (a), (b), (c) and (d), 
the observed values of the field $\chi$ (in MeV) with (without) AMM  
are 406.57 (406.49), 406.574 (406.6), 406.58 (406.594)
and 407.12 (406.77) respectively.
For density, $\rho_B=5\rho_0$ in symmetric nuclear matter,
for the same magnetic fields, these values of $\chi$
are modified to 396.38 (395.6), 397.2 (394.57), 397.7 (394.17), 
398.29 (393.88) respectively. 
As the isospin asymmetry of the medium increases, 
the scalar field $\chi$ is also observed to increase.
At density of $5\rho_0$, and for magnetic field $|eB|=4 m_\pi^2$, 
compared to the value of 396.38 (395.6) for symmetric nuclear matter,
the scalar field $\chi$ has the values to be  396.51 (396) 
and 397.64 (397.34) MeV for isospin asymmetry 
parameter $\eta$ as 0.3 and 0.5 respectively, 
with (without) accounting for the
anomalous magnetic moment effects for the nucleons.  
For the same density, and for the value of the magnetic 
field as $|eB|=12 m_\pi^2$, 
the value of $\chi$ is modified from 
398.29 (393.88) for symmetric case, 
to 398.42 (395.61) and 399.84 (397.34) MeV, 
for $\eta$ as 0.3 and 0.5 respectively, 
with (without) AMM effects.
For the case of $\eta$=0.5, the medium comprises of
only neutrons, and hence the only effect of magnetic field
is due to the anomalous magnetic moment of the neutrons.
Hence, in the case when the AMM efects are not taken into
consideration, the value of the dilaton field remains
independent of the magnetic field. 
This leads to the mass shift of the charmonium states
to be independent of the magnetic field, when AMM
effects are not taken into account.

We compute the shifts of the masses of the charmonium states, 
$J/\psi$, $\psi(3686)$ and $\psi(3770)$, from their
vacuum values, as arising from the medium
change of the dilaton field $\chi$ in the chiral SU(3) model.
These are plotted in
figures \ref{mjpsi_mag}, \ref{mpsi_2s_mag} and \ref{mpsi_1d_mag}, 
as functions of baryon density (in units of nuclear matter
saturation density). In the massless QCD limit,
the mass shift is observed to be proportional to the medium change
in the fourth power of the $\chi$ field, as can be seen from
equation (\ref{masspsi}). 
for given values of the magnetic field, and for different values of the
isospin asymmetry parameter, $\eta$. 
We have shown the results for the 
values of the magnetic field as $|eB|$, 
as $4m_\pi^2$, 
 $8m_\pi^2$,  $10m_\pi^2$,  $12m_\pi^2$ 
in panels (a),
(b), (c) and (d) for values of isospin asymmetry
parameter, $\eta$ as 0, 0.3 and 0.5. The effects of 
anomalous magnetic moments on the in-medium charmonium
masses are also investigated and compared to the cases
when these effects are not taken into account.

The change in the mass of $J/\psi$ as a function of nuclear matter density 
is observed to be relatively small as compared to that of $\psi (3686)$ and
$\psi(3770)$. There is a significant drop in the  masses of $\psi(3686)$
and $\psi(3770)$ as a function of density. The drop in these masses are
the largest for isospin asymmetry parameter $\eta=0$ in both the cases 
of accounting for AMM and without AMM effects. 
The mass shifts of all these charmonium states from the vacuum values,
are observed to decrease, when the isospin asymmetry of the medium 
is increased. 
There is observed to be comparatively less drop in the masses of
charmonium states when we account for the calculations with AMM 
as compared to the case of without AMM calculations.
The difference in the mass shift of charmonium states 
accounting the AMM and without accounting
the AMM increases as we increase the magnetic field. 

For symmetric nuclear matter ($\eta$=0), 
at $\rho_B=\rho_0 (5\rho_0$), including the effects of AMM, the mass
shift (in MeV) of $J/\psi$ is obtained as $-8.16$ $(-32.93)$, 
$-8.16$ $(-31)$, $-8.13$ $(-29.8)$ and $-8.22$ $(-28.44)$ for $|eB|$
as $4m_\pi^2$, $8m_\pi^2$, $10m_\pi^2$, and, $12m_\pi^2$ 
respectively. When the AMM effects are not accounted for,
these values are observed to be
$-8.37$ $(-34.745)$, $-8.42$ $(-37.14)$, $-8.45$ $(-38.05)$ 
and $-8.5$ $(-38.73) $ MeV respectively.
The value of mass shift of $J/\psi$ at nuclear matter density, 
$\rho_0$, in the presence of magnetic field, may be compared to
the value of around $-8.6$ MeV in the absence of magnetic field 
\cite{amarvdmesonTprc} 
in the chiral effective model as used in the present investigation.
As has already been mentioned the in-medium masses of the
charmonium states are computed from the medium changes of
the gluon condensates calculated from the modifications
of the dilaton field within the chiral effective model and
there is no assumption of linear density approximation.
The results of the present work are thus not restricted to 
low densities, but are also valid for high densities 
as well.
The value of mass shift of $J/\psi$ of around $-8.6$ MeV,
at $\rho_B=\rho_0$ for zero magnetic field,
is similar to the value ($\sim -8 $MeV) obtained using the 
QCD Stark effect, where the change in the
scalar gluon condensate in the nuclear medium 
is due to the change in the expectation value of 
$\langle \frac{\alpha_s}{\pi} {\vec E}^2 \rangle$,
computed using the linear density approximation
\cite{leeko}.  
In the absence of the magnetic field, 
at the nuclear matter saturation density, 
the values of the mass shifts of the excited charmonium states
$\psi (3686)$ and $\psi(3770)$, calculated within the
chiral effective model \cite {amarvdmesonTprc} are 
$-117$ and $-155$ MeV respectively in
nuclear matter. These values may be compared 
with the values of the mass shifts of $-100$ and $-140$ MeV,
for these charmonium states, as calculated using the 
linear density approximation \cite{leeko}.

The medium modifications of the masses of the charmonium states 
given by equation (\ref{masspsi}) are due to the medium
modification of the gluon condensate, calculated within 
the effective chiral model by the medium change of the 
dilaton field. The wave function of the charmonium state,
for the harmonic oscillator potential is given by
equation (\ref{wavefn}). Applying the formula to the case of $\eta_c$, 
which is also a 1S state, we observe 
a similar trend of the mass shift of $\eta_c$ as for $J/\psi$.
We calculate the harmonic oscillator potential strength,
$\beta$ for $\eta_c$ meson, to be 535.2 MeV, 
by linear extrapolation from the vacuum mass versus $\beta$
graph, for $J/\psi$ and $\psi(3686)$. 
In symmetric nuclear matter,
for density, $\rho_B=\rho_0 (5\rho_0$),
the mass shift of $\eta_c$ is observed to be 
$ −6.87$ $(−26.98)$, $−6.77$ $(−25.36)$, $−6.77$ $(−24.30)$ and 
$−6.77$ $(−23.33) $ MeV for $|eB|$ 
as $4m_\pi^2$, $8m_\pi^2$, $10m_\pi^2$, and, $12m_\pi^2$ 
respectively, when AMM effects are taken into account.
These values are modified to 
$−6.89$ $(−28.39)$, $−6.93$ $(−30.39)$, $−6.95$ $(−31.15)$ 
and $−6.99$ $(−31.61)$ MeV respectively, when the AMMs of the nucleons
are not taken into consideration.
In the absence of magnetic field, the mass shift of
$\eta_c$ meson is observed to be $-7.1$ ($-27.65$)  MeV 
for density of $\rho_0 (5\rho_0)$. The value at
nuclear matter density may be compared to the 
value of $-5.69$ MeV calculated using the QCD sum rule approach
\cite{amarvjpsi_qsr} as well as to the value of
$-3$ MeV calculated from the $\eta_c$--nucleon scattering
length in Ref. \cite{kaidalov_etac_mass}
similar to the calculations in Ref. \cite{pes2}
for $J/\psi$ nucleon scattering amplitude.

For the value of isospin asymmetry parameter $\eta$ as 0.3, 
at $\rho_B=\rho_0 (5\rho_0)$,
when the AMM effects are not accounted for,
the mass shift (in MeV) of $\psi (3686)$ is
$-109.8$ $(-458.72)$, $-109.88$ $(-468.42)$, 
$-109.88$ $(-470.13)$ and $-109.88$ $(-471.14)$
for value of the magnetic field 
$|eB|$ as $4m_\pi^2$, $8m_\pi^2$, 
 $10m_\pi^2$, and, $12m_\pi^2$ respectively. 
These values are modified to $-106.38$ $(-442.47)$,
$-106.38$ $(-421.2)$, $-102.72$ $(-401.63)$ and $-102.95$ $(-381.41)$,
when the AMM effects are taken into account.
In the same conditions ($\eta$=0.3 and density of $\rho_0 (5\rho_0)$), 
the mass shift in $\psi (3770)$ (in MeV) 
is $-145.13$ $(-606.3)$, $-145.23$ $(-619.14)$, $-145.23$ $(-621.4)$
and $-145.23$ $(-622.73)$ respectively 
for $|eB|$ as $4m_\pi^2$, $8m_\pi^2$, $10m_\pi^2$, and, $12m_\pi^2$ 
when the AMM effects are not accounted for,
and, $-140.6$ $(-584.84)$, 
$-140.6$ $(-556.72)$, $-135.77$ $(-530.86)$ and $-136.07$ $(-504.13)$,
when the AMM effects are taken into account.

For the maximum isospin asymmetry in the system ($\eta=0.5$),
which corresponds to neutron matter, 
at $\rho_B=\rho_0 (5\rho_0)$, without accounting 
for the effects of AMM, the mass shifts  (in MeV)
in $J/\psi$, $\psi (3686)$ and $\psi (3770)$
are observed to be $-7.86(-30.68)$, $-106.61(-416.13)$ 
and $-140.92(-550.03)$ MeV
respectively, and are independent of the magnetic field.
As has alredy been mentioned, this is because there are 
no charged fermions (protons) in the system, and, 
the neutrons interact with magnetic field
only due to their AMM.
When the AMM effects are taken into account, for $\eta$=0.5,
for density $\rho_B=\rho_0 (5\rho_0)$, the mass shifts for the 
$J/\psi$ are $-7.36$ $(-29.96)$, $-7.24$ $(-27.83)$,
$-7.17$ $(-26.31)$ and $-7.09$ $(-24.73)$
for the value of the magnetic field 
$|eB|$ as $4m_\pi^2$, $8m_\pi^2$, 
 $10m_\pi^2$, and, $12m_\pi^2$ repectively. 
For $\psi (3686)$, for the same conditions, these are modified to 
$-99.8$ $(-406.48)$, $-98.24$ $(-377.56)$, $-97.31$ $(-356.89)$
and $-96.25$ $(-335.45)$ and, for $\psi (3770)$, to 
$-131.91$ $(-537.26)$, $-129.85$ $(-499)$, $-128.62$ $(-471.72)$
and $-127.21$ $(-443.4)$ respectively.

\section{Summary}

To summarize, we have investigated the effects of density, 
isospin asymmetry, magnetic field and AMM of nucleons 
on the mass modifications of the charmonium states $J/\psi (3097)$, 
$\psi(3686)$ and $\psi (3770)$ from the modification of the scalar 
dilaton field in isospin asymmetric nuclear medium in presence 
of strong magnetic fields using a SU(3) chiral model. 
The excited charmonium states are observed to show
significant modifications in their masses in the medium,
as compared to the mass of $J/\psi$ in the nuclear medium.
The value of the mass shift of the $J/\psi$ as well as 
other charmonium states are observed to increase with density.
For a given value of density and magnetic field, the effect of the 
isospin asymmetry of the medium is to increase the masses of the 
charmonium states. The effect 
of the magnetic field on the charmonium states is found 
to decrease the values of mass shift when the AMM effects are 
taken into account, as compared to when these are not
taken into consideration.
The mass shifts of the charmonium states in the
nuclear medium in presence of magnetic fields,
seem to be appreciable at high densities and these should show in
observables like the production of these charmonium states, as well as
of the open charmed mesons in the compressed baryonic matter (CBM)
experiment at the future facility at GSI, where baryonic matter 
at high densities and moderate temperatures will be produced.

\acknowledgements
One of the authors (AM) is grateful to ITP, University of Frankfurt,
for warm hospitality and 
acknowledges financial support from Alexander von Humboldt Stiftung 
when this work was initiated. 
Amal Jahan CS acknowledges the support
towards this work from Department of Science and Technology, Govt of India,
via INSPIRE fellowship scheme  offer (Ref. No. DST/INSPIRE/03/2016/003555).



\begin{thebibliography}{}
\bibitem{Hosaka_Prog_Part_Nucl_Phys} 
A. Hosaka, T. Hyodo, K. Sudoh, Y. Yamaguchi, S. Yasui,
Prog. Part. Nucl. Phys. {\bf 96}, 88 (2017).
\bibitem{HIC_mag} V. Skokov, A. Y. Illarionov and V. Toneev,
Int. J. Mod. Phys. A {bf 24}, 5925 (2009); W. T. Deng and X.G.Huang,
Phys.Rev. C {\bf 85}, 044907 (2012); D. Kharzeev, L. McLerran and 
H. Warringa, Nucl. Phys. A {\bf 803}, 227 (2008); K. Fukushima,
D. E. Kharzeev and H. J. Warringa, Phys. Rev. D {\bf 78}, 074033
(2008).

\bibitem{time_evolution_B_HIC_Tuchin}
K. Tuchin, Phys. Rev. C {\bf 83}, 017901 (2011);
K. Marasinghe and K. Tuchin, Phys. Rev. C {\bf 84},
044908 (2011);
K. Tuchin, Phys. Rev. C {\bf 83}, 034904 (2010),
K. Tuchin, Erratum Phys. Rev. C {\bf 83}, 039903(E) (2011);
K. Tuchin, Phys. Rev. C {\bf 88}, 024911 (2013).
\bibitem{time_evolution_B_HIC_Ajit}
Arpan Das, S. S. Dave. P.S. Saumia and A.M. Srivastava, 
Phys. Rev. C {\bf 96}, 034902 (2017).

\bibitem{Gubler_D_mag_QSR} P. Gubler, K. Hattori, S. H. Lee, M. Oka,
S. Ozaki and K. Suzuki, Phys. Rev. D {\bf 93}, 054026 (2016).

\bibitem{machado_1} C. S. Machado, F. S. Navarra. E. G. de Oliveira
and J. Noronha, Phys. Rev. D {\bf 88}, 034009 (2013).

\bibitem{B_mag_QSR} C. S. Machado, R.D. Matheus, S.I. Finazzo and
J. Noronha, Phys. Rev. D {\bf 89}, 074027 (2014).

\bibitem{dmeson_mag}
Sushruth Reddy P, Amal Jahan CS, Nikhil Dhale, Amruta Mishra,
J. Schaffner-Bielich, Phys. Rev. C {\bf 97}, 065208 (2018).
\bibitem{bmeson_mag}
Nikhil Dhale, Sushruth Reddy P, Amal Jahan CS, Amruta Mishra,
Phys. Rev. C {\bf 98}, 015202 (2018).

\bibitem{charmonium_mag_QSR} S. Cho, K. Hattori, S. H. Lee, K. Morita
and S. Ozaki, Phys. Rev. Lett. {\bf 113}, 122301 (2014).
\bibitem{charmonium_mag_lee} 
S. Cho, K. Hattori, S. H. Lee, K. Morita
and S. Ozaki, Phys. Rev. D {\bf 91}, 045025 (2015).

\bibitem{eichten}
E.Eichten, K. Gottfried, T. Kinoshita, K.D. Lane and
T.M. Yan, Phys. Rev. D {\bf 17}, 3090 (1978);
ibid, Phys. Rev. D {\bf 21}, 203 (1980).
\bibitem{satz}
L. Kluberg and H. Satz, arXiv:0901.3831 (hep-ph);
F. Karsch, M. T. Mehr, and H. Satz, 
Z. Phys. C37, 617 (1988);
A. Bazavov, P. Petreczky, and A. Velytsky, 
arXiv: 0904.1748 (hep-ph);
S. Digal, P. Petreczky, and H. Satz, 
Phys. Lett. B514, 57 (2001); 
A. Mocsy and P. Petreczky, 
Phys.Rev.  D73, 074007 (2006), hep-ph/0512156.
\bibitem{repko}
S.F. Radford and W W. Repko, Phys.Rev D {\bf 75}, 074031 (2007).
\bibitem{pes1} M.E. Peskin, Nucl. Phys. {\bf B156}, 365 (1979).
\bibitem{pes2} 
G. Bhanot and M.E. Peskin, Nucl. Phys. {\bf B156}, 391 (1979).
\bibitem{voloshin}
M.B.Voloshin, Nucl. Phys. B154 ,365 (1979).

\bibitem{kimlee}
Sugsik Kim, Su Houng Lee, Nucl. Phys. A {\bf 679}, 517 (2001).
\bibitem{klingl} F. Klingl $et al.$, Phys. Rev. Lett. {\bf 82}, 3396 (1999).
\bibitem{amarvjpsi_qsr} 
Arvind Kumar and Amruta Mishra, Phys. Rev. C {\bf 95}, 065206 (2010).
\bibitem{hatsuda}  T. Hatsuda, S.H. Lee, Phys. Rev. C {\bf 46}, R34, (1992).
\bibitem{am_vecmeson_qsr} Amruta Mishra, Phys. Rev. C {\bf 91} 035201
(2015).
\bibitem{open_heavy_flavour_qsr} 
Arata Hayashigaki , Phys. Lett. B {\bf 487}, 96 (2000);
T. Hilger, R. Thomas and B. K\"ampfer, Phys. Rev. C {bf 79},
025202 (2009); T. Hilger, B. K\"ampfer and S. Leupold, 
Phys. Rev. C {\bf 84}, 045202 (2011); 
S. Zschocke, T. Hilger and B. K\"ampfer,
Eur. Phys. J. A {\bf 47} 151 (2011).

\bibitem{Wang_heavy_mesons} 
Z-G. Wang and Tao Huang, Phys. Rev. C {\bf 84}, 048201 (2011);
Z-G. Wang, Phys. Rev. C {\bf 92}, 065205 (2015).

\bibitem{arvind_heavy_mesons_QSR}
Rahul Chhabra and Arvind Kumar, Eur. Phys. J A
{\bf 53}, 105 (2017); ibid, Eur. Phys. J C {\bf 77},
726 (2017), Arvind Kumar and Rahul Chhabra, 
Phys. Rev. C {\bf 92}, 035208 (2015).

\bibitem{open_heavy_flavour_qmc}
	K. Tsushima, D. H. Lu, A. W. Thomas, K. Saito, and R. H. Landau,
	Phys. Rev. C {\bf 59}, 2824 (1999);
	A. Sibirtsev,	K. Tsushima, and A. W. Thomas,
	Eur. Phys. J. A {\bf 6}, 351 (1999);
        K. Tsushima and F. C. Khanna, Phys. Lett. B {\bf 552}, 138
        (2003).
\bibitem{qmc} P. A. M. Guichon, Phys. Lett. B {\bf 200}, 235
(1988); K. Saito and A. W. Thomas, Phys. Lett B {\bf 327}, 9 (1994);
K. Saito, K. Tsushima and A. W. Thomas, Nucl. Phys. A {\bf 609}, 339
(1996); P.K. Panda, A. Mishra, J. M. Eisenberg and W. Greiner,
Phys. Rev. C {\bf 56}, 3134 (1997).

\bibitem{Yasui_Sudoh_pion}
S. Yasui and K. Sudoh, Phys. Rev. C {\bf 87}, 015202 (2013).
\bibitem{Yasui_Sudoh_heavy_meson_Eff_th}
S. Yasui and K. Sudoh, Phys. Rev. C {\bf 89}, 015201 (2014).

\bibitem{Yasui_Sudoh_heavy_particle_impurity}
S. Yasui and K. Sudoh, Phys. Rev. C {\bf 88}, 015201 (2013).

\bibitem{ltolos} L.Tolos, J. Schaffner-Bielich and A. Mishra,
Phys. Rev. {\bf C 70}, 025203 (2004).
\bibitem{ljhs} L. Tolos, J. Schaffner-Bielich
and H. St\"ocker, Phys. Lett. {\bf B 635}, 85 (2006).
\bibitem{mizutani} T. Mizutani and A. Ramos, Phys. Rev. {\bf C 74},
065201 (2006); L.Tolos, A. Ramos and T. Mizutani, Phys. Rev. {\bf C 77},
015207 (2008).
\bibitem{HL}J. Hofmann and M.F.M.Lutz, Nucl. Phys. {\bf A 763}, 90 (2005).

\bibitem{tolos_heavy_mesons}
R. Molina, D. Gamermann, E. Oset, and L. Tolos,
Eur. Phys. J A {\bf 42}, 31 (2009);
L. Tolos, R. Molina, D. Gamermann, and E. Oset, 
Nucl. Phys. A {\bf 827} 249c (2009).

\bibitem{leeko} Su Houng Lee and Che Ming Ko, 
Phys. Rev. C {\bf 67}, 038202 (2003).
\bibitem{moritalee}
K. Morita and S.H. Lee, Phys. Rev. C {\bf 77}, 064904 (2008);
S.H. Lee and K. Morita, Phys. Rev. D {\bf 79}, 011501(R) (2009);
K. Morita and S.H. Lee, Phys. Rev. C {\bf 85}, 044917 (2012);
K. Morita and S.H. Lee, Phys. Rev. Lett {\bf 100}, 022301 (2008).

\bibitem{Schechter} J. Schechter, Phys. Rev. D {\bf 21}, 3393 (1980).
\bibitem{paper3}
 	P. Papazoglou, D. Zschiesche, S. Schramm, J. Schaffner-Bielich,
	H. St\"ocker, and W. Greiner, Phys. Rev. C {\bf 59},  411  (1999).
\bibitem{kristof1}
	A. Mishra, K. Balazs, D. Zschiesche, S. Schramm,
	H. St\"ocker, and W. Greiner,
        Phys. Rev. C {\bf 69}, 024903 (2004). 
\bibitem{amarindamprc}
Amruta Mishra and Arindam Mazumdar, Phys. Rev. C {\bf 79},  024908 (2009). 
\bibitem{amarvdmesonTprc}
Arvind Kumar and Amruta Mishra, Phys. Rev. C {\bf 81}, 065204
(2010).
\bibitem{amarvepja}
Arvind Kumar and Amruta Mishra, Eur. Phys. A {\bf 47}, 164
(2011).
\bibitem{krein_jpsi}
G. Krein, A. W. Thomas and K. Tsushima, Phys. Lett. B {\bf 697},
136 (2011).
\bibitem{krein_17}
G. Krein, A. W. Thomas and K. Tsushima, arXiv: 1706.02688
(hep-ph).
\bibitem{hartree}
	D. Zschiesche, A. Mishra, S. Schramm, H. St\"ocker and W. Greiner,
        Phys. Rev. C {\bf 70}, 045202 (2004).
\bibitem{kaon_antikaon}
A. Mishra, E. L. Bratkovskaya, J. Schaffner-Bielich, S. Schramm
     and H. St\"ocker, Phys. Rev. C {\bf 70}, 044904 (2004).
\bibitem{isoamss}
A. Mishra and S. Schramm, Phys. Rev. C {\bf 74}, 064904 (2006).	
\bibitem{isoamss1}
A. Mishra, S. Schramm and W. Greiner, Phys. Rev. C {\bf 78}, 
024901 (2008).
\bibitem{isoamss2}
Amruta Mishra, Arvind Kumar, Sambuddha Sanyal, S. Schramm,
Eur. Phys, J. A {\bf 41}, 205  (2009).  
\bibitem{pneutronstar}
Amruta Mishra, Arvind Kumar, Sambuddha Sanyal, V. Dexheimer, 
S. Schramm, Eur. Phys. J {\bf 45}, 169 (2010).
\bibitem {amdmeson} 
A. Mishra, E. L. Bratkovskaya, J. Schaffner-Bielich, 
S.Schramm and H. St\"ocker, Phys. Rev. {\bf C 69}, 015202 (2004).
\bibitem{DP_AM_Ds}
Divakar Pathak and Amruta Mishra, Adv. High Energy Phys. 2015,
697514 (2015).
\bibitem{DP_AM_bbar}
Divakar Pathak and Amruta Mishra, Phys. Rev. C {\bf 91}, 045206
(2015).
\bibitem{DP_AM_Bs}
Divakar Pathak and Amruta Mishra, Int. J. Mod. Phy. E {\bf 23}, 
1450073 (2014).
\bibitem{AM_DP_upsilon}
Amruta Mishra and Divakar Pathak, Phys. Rev. C {\bf 90}, 025201
(2014).
\bibitem{friman}
B. Friman, S. H. Lee and T. Song, Phys. Lett, B {\bf 548}, 153
\bibitem{3p0}
A. Le Yaouanc, L. Oliver, O. Pene and  J. C. Raynal, Phys. Rev. D
{\bf 8}, 2223 (1973); ibid, Phys. Rev. D {\bf 9}, 1415 (1974);
ibid, Phys. Rev. D {\bf 11}, 1272 (1975);
T. Barnes, F. E. Close, P. R. Page and E. S. Swanson, Phys. Rev. D
{\bf 55}, 4157 (1997).
(2002).
\bibitem{amspmwg}
Amruta Mishra, S. P. Misra and W. Greiner, Int. J. Mod. Phys.
E {\bf 24}, 155053 (2015).
\bibitem{amspm_upsilon}
Amruta Mishra and S. P. Misra, Phy. Rev. C {\bf 95}, 065206 (2017).
\bibitem{weinberg}
S.Weinberg, Phys. Rev. {\bf 166} 1568 (1968).
\bibitem{coleman}
S. Coleman, J. Wess, B. Zumino, Phys. Rev. {\bf 177} 2239 (1969);
C.G. Callan, S. Coleman, J. Wess, B. Zumino, Phys. Rev. {\bf 177}
2247 (1969).
\bibitem{bardeen}
W. A. Bardeen and B. W. Lee, Phys. Rev. {\bf 177} 2389 (1969).
\bibitem{heide1}
Erik K. Heide, Serge Rudaz and Paul J. Ellis, Nucl. Phys. A 
{\bf 571}, (2001) 713.
\bibitem{broderick1}
A. Broderick, M. Prakash and J.M.Lattimer, Astrophys. J. 537, 351
(2002). 
\bibitem{broderick2}
A.E. Broderick, M. Prakash and J. M. Lattimer, Phys. Lett. {\bf B}
531, 167 (2002).
\bibitem{Wei}
F. X. Wei, G. J. Mao, C. M. Ko, L. S. Kisslinger, H. St\"ocker,
and W. Greiner, J. Phys. G, Nucl. Part. Phys. {\bf 32},
47 (2006).
\bibitem{mao} Guang-Jun Mao, Akira Iwamoto, Zhu-Xia Li,
Chin. J. Astrophys. 3, 359 (2003).
\bibitem{amm}
M. Pitschmann and A. N. Ivanov, arXive : 1205.5501 (math-ph). 

\bibitem{VD_SS}
V. Dexheimer, R. Negreiros, S. Schramm, Eur. Phys. Journal A 
{\bf 48}, 189 (2012); V. Dexheimer, B. Franzon and S. Schramm,
Jour. Phys. Conf. Ser. {\bf 861}, 012012 (2017).
\bibitem{aguirre_fermion}
R. M. Aguirre and A. L. De Paoli, Eur. Phys. J. A {\bf 52}, 343
(2016).

\bibitem{cohen}
Thomas D. Cohen, R. J. Furnstahl and David K. Griegel,
 Phys. Rev. C {\bf 45}, 1881 (1992).
\bibitem{kaidalov_etac_mass}
A. B. Kaidalov, P.E. Volkovitsky, Phys. Rev. Lett. {\bf 69},
3155 (1992).
\end{thebibliography}
\end{document}